# Coherent Spin Pumping Originated from Sub-Terahertz Néel Vector Dynamics in Easy Plane α-Fe$_2$O$_3$ /Pt

Gregory Fritjofson[1], Atul Regmi[1], Jacob Hanson-Flores[1], Justin Michel[2], Junyu Tang[3], Fengyuan Yang[2], Ran Cheng[4,3], Enrique Del Barco[1]*

[1] Department of Physics, University of Central Florida, Orlando, FL 32765, USA
[2] Department of Physics, The Ohio State University, Columbus, Ohio 43210, USA
[3] Department of Physics and Astronomy, University of California, Riverside, California 92521, USA
[4] Department of Electrical and Computer Engineering, University of California, Riverside, California 92521, USA

*Corresponding author email: [delbarco@ucf.edu]

## Abstract

We present a thorough study of spin-to-charge current interconversion in bulk and thin films of (0001) α-Fe$_2$O$_3$ /Pt heterostructures by means of all-optical polarization-controlled microwave excitation at sub-Terahertz frequencies. Our results demonstrate that coherent spin pumping is generated through excitations of both the acoustic and optical modes of antiferromagnetic resonance, provided that the corresponding selection rules are met for the relative orientation between the microwave magnetic field $\vec{h}_{ac}$ and the magnetic moment $\vec{m}_0$ of the Hematite. In particular, our results unanimously show that while a microwave field with $\vec{h}_{ac} \perp \vec{m}_0$ pumps a net spin angular momentum from the acoustic mode, spin pumping from the optical mode is only enabled when $\vec{h}_{ac} \parallel \vec{m}_0$, as expected from the selection rules imposed by the Néel vector dynamics. Our results support the current understanding of spin mixing conductance in antiferromagnetic/non-magnetic interfaces, contrary to recent reports where the absence of spin pumping from the optical mode in Hematite was interpreted as a cancellation effect between the diagonal and off-diagonal components of the spin mixing conductance. We also provide an explanation for the previously reported observations and show how the optical spin pumping actually vanishes for thin films, which we speculate being either due to an increased level of inhomogeneities or to insufficient film thickness for the optical mode to fully realize.

**Main**

Recent technological advances in spintronic devices have motivated a growing research landscape in magnetic materials for their use in spintronics applications, such as spin valves, racetrack/magnetic memory, or spin logic among others.[1–3] Antiferromagnets (AFMs) featuring high frequency dynamics are gaining special attention for their potential in next-generation THz device applications, and their abundance, durability, tunability, robustness, and unique juxtaposition of multiferroics and spin-stability[1,4,5]. AFM spintronics has witnessed substantial progress in recent years, highlighted by the demonstration of coherent spin pumping and the ensuing spin-charge interconversion at AFM/NM interfaces, carving the path for direct access to the high frequency dynamics of the Néel order[6,7]. Meanwhile, imaging and read-out of the AFM order parameter[8,9] along with ultrafast studies[10–13] of AFMs have significantly developed, and real-world applications of AFM memory[14] are on the rise. Nevertheless, achieving high-frequency coherent spintronics in real-world on-chip devices calls for advances towards miniaturization and practicality.

The spin-pumping effect realized in antiferromagnets fluorides ($MnF_2$)[6] and oxides ($Cr_2O_3$)[7] has sparked surging interest in insulating AFM-based devices harnessing THz frequency dynamics, where the use of spin-pumping and the inverse-spin-Hall-effect (ISHE) presents a relatively simple scenario to generate and detect THz signals[15,16]. In such a device, an AFM is capped with a non-magnetic (NM) conductive layer with a large spin-orbit coupling, *e.g.* a heavy metal such as Pt. In pure AFM spin pumping, the Néel order is coherently excited by incident microwaves in the form of antiferromagnetic resonance (AFMR). Under these conditions, angular momentum is transferred into the adjacent NM layer and characterized by the spin-mixing conductance of the AFM/NM interface. The result is a polarized spin current traversing the interface, converting into a measurable ISHE voltage in the NM layer.

While coherent spin pumping has been experimentally demonstrated in easy-axis AFMs[6,7] as well as easy-plane AFMs[17,18], confirming the theoretical picture[19], recent studies suggested the existence of cross-sublattice contributions to the spin-mixing conductance[20], which could lead to partial cancellation of spin pumping, bringing the commonly accepted understanding into question. Hattori *et al*.[21] have recently reported a null spin pumping from the high-frequency (optical) mode while corroborating the effect from the low-frequency (acoustic) mode, which they attributed to the cancellation between intra- and cross-sublattice terms in the spin-mixing conductance, supporting recent theoretical speculations[22]. If proven right, its implications would have tremendous repercussions as it would allegedly preclude the use of a particular class of AFM systems (e.g., easy-plane AFMs) for THz spintronics applications. However, a close inspection of this experiment reveals a controversial issue. As the Poynting vector of the microwave is parallel to the magnetic moment $\vec{m}_0$, the oscillating magnetic field $\vec{h}_{ac}$ cannot properly couple, hence drive, the linearly polarized magnetization in the optical mode, making it ambiguous to claim the absence of spin pumping out of the optical mode.

Here we study coherent spin-pumping in the same heterostructure, α-Fe$_2$O$_3$/Pt, designed to address this critical open question. We have performed experiments in both bulk and thin-film α-Fe$_2$O$_3$ samples, placing special attention to the α-Fe$_2$O$_3$ magneto-crystalline alignment with respect to the microwave stimuli of different frequencies and polarizations. Our results demonstrate that spin pumping can indeed be generated by exciting both the acoustic and optical modes of Hematite, supporting the current understanding of coherent spin pumping in AFM/NM interfaces without the cross-sublattice components.[19] Our findings also provide the first demonstration of coherent spin pumping from the sub-THz dynamics associated with the Néel order parameter in an easy-plane AFM system, timely expanding the recent discovery of its counterpart associated with the small magnetization in the same system[17,18].

In Hematite, α-Fe$_2$O$_3$, a naturally abundant and robust AFM with a high Néel temperature of 950 K, the Dzyaloshinskii-Moriya interaction (DMI) produces a non-zero magnetization in the basal plane above the Morin temperature. In our experiments, we choose (0001)-oriented α-Fe$_2$O$_3$ bulk and thin film samples with the device configuration shown in Figure 1a. Bulk α-Fe$_2$O$_3$(0001) was purchased from MTI and capped in-situ with a 5nm Pt layer. The capping was performed at 4 mTorr in Ar plasma, at 20mA with DC off-axis sputtering and a base pressure below 1e-7 Torr. The α-Fe$_2$O$_3$(0001) film was grown epitaxially on an Al$_2$O$_3$(0001) substrate by off-axis sputtering using a 50 W RF power with a substrate temperature of 400 °C and 12 mTorr pressure of Ar + 5% O$_2$. The film was capped in-situ with a 5 nm Pt layer grown with DC off-axis sputtering in 2 mTorr Ar at room temperature. Contact pads of Cr(7.5nm)/Au(50nm) were deposited by PVD for both samples. In this crystal orientation, the Néel vector lies primarily within the sample plane, and magnetic excitations produce spin polarization parallel to the sample surface, as needed for the eventual conversion into a measurable transverse ISHE voltage in the adjacent NM layer.

In our experiments, the spin-to-charge conversion is induced by sub-THz optical microwaves, where the polarization state is precisely controlled via a Martin-Puplett Interferometer. Linear microwave polarizations, referring to the direction of the ac magnetic field $\vec{h}_{ac}$, are used for all experiments (see Fig. 1a). A dc magnetic field is applied in the plane of the film so that $\vec{H} \perp \vec{c}(\vec{z})$, while the microwave may be applied using either the Faraday geometry where it propagates in the $\vec{x}$-direction (along $\vec{H}$) or the Voigt geometry where it propagates in the $\vec{z}$-direction (perpendicular to $\vec{H}$). The microwaves propagate entirely in free-space, and the cross-polarized reflected component is collected. The details of our experimental setup have been reported previously[23]. The ISHE circuit is enclosed in an optical cryostat and held in the center of a split-bore 9 T superconducting magnet. All measurements are performed at room temperature (for temperature dependence see Fig. S4).

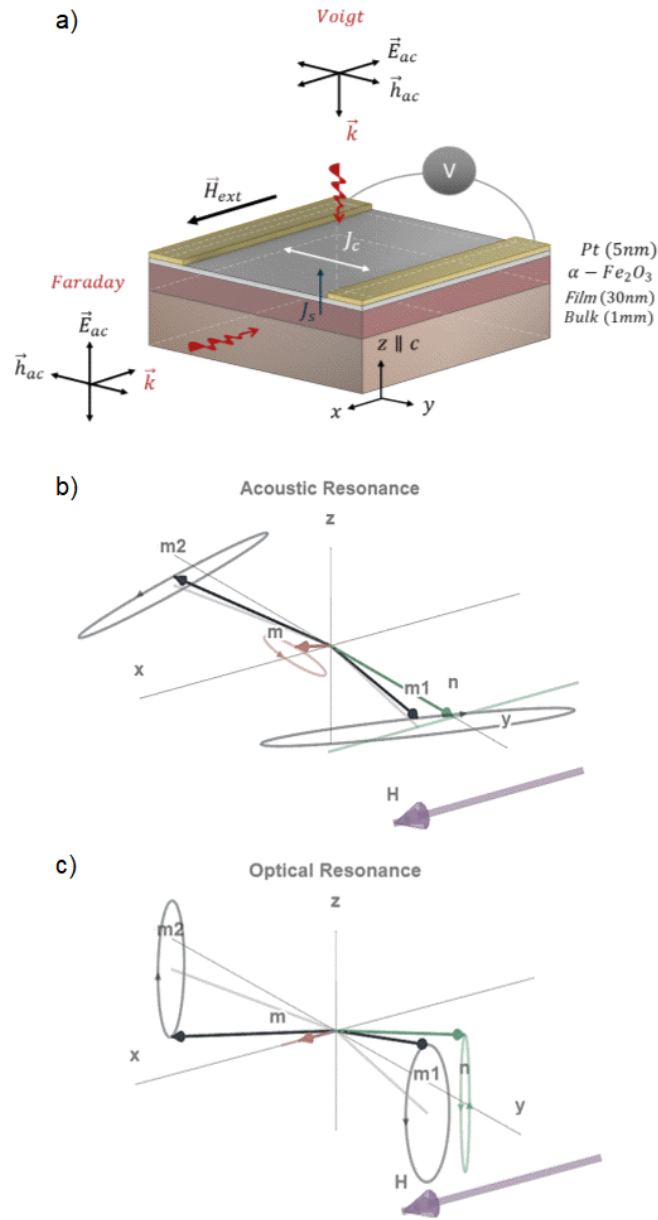

**Figure 1:** (a) ISHE device showing optical microwave excitation and process of spin-to-charge conversion within α-Fe2O3/Pt. Variations of microwave propagation and linear polarizations direct $\vec{h}_{ac}$ along the different crystal principal axes. (b-c) Graphical illustrations of the magnetization dynamics for the acoustic and optical mode, respectively. The canting angle of the magnetizations and the strength of the DMI have been exaggerated in the figure for illustration purposes.

Contrary to easy-axis AFMs such as MnF$_2$, easy-plane antiferromagnets like α-Fe$_2$O$_3$ possess optical and acoustic resonance modes that are well separated in frequency (by over 100 GHz) in the absence of an external field. The dynamics for both modes are schematically illustrated in Figures 1b-c. In the acoustic mode (Figure 1b), the net magnetization precesses elliptically around its equilibrium (along the x-direction) thanks to the DMI, while the Néel vector undergoes a strictly linear oscillation confined to the basal plane (x-y plane). Therefore, the ISHE signal collected from the excitation of the acoustic mode is related to the dynamics of the net magnetization (a quasi-ferromagnetic mode), which would be absent without the DMI. In the optical mode, the net magnetization vibrates linearly along the $\vec{x}$-axis so that it cannot generate any dc spin pumping. Therefore, a non-zero ISHE signal resulting from the excitation of the optical mode would be a direct evidence of spin pumping from the Néel vector. From a theoretical perspective, even though the Néel vector undergoes a linear oscillation in the y-z plane, its magnitude also varies periodically, thus leading to a net dc spin pumping from the optical mode.

We begin by investigating the response of our bulk Hematite heterostructure sample, α-$Fe_2O_3$(0.5 mm)/Pt(5 nm). Figure 2a shows the AFMR (upper spectra) and ISHE voltages (lower spectra) for traces at 160 GHz for field sweeps up to 5.5 T in the Voight geometry (i.e., microwaves directed parallel to the applied dc magnetic field, $\vec{k} \parallel \vec{z}$,). Low-field (high-field) microwave absorptions correspond to the optical (acoustic) branch. In this arrangement, a linearly polarized light along the principal axes could induce the resonance of either mode (upper spectra in Fig. 2a). However, the corresponding ISHE voltages are selective for these modes: for the acoustic (optical) branch, spin-to-charge transfer is only achieved for y-polarized (x-polarized) light, and this feature persists for all frequencies measured. We note that the observation of a spin pumping signal from the optical mode in our sample is made possible by the required orientations between $\vec{h}_{ac}$ and the Néel vector, which was not properly chosen in the experiments conducted by Hattiro et al.[21], possibly explaining their null observation.

As shown in Figure 2b, the eigen-frequencies for both modes are fit to[24,25]:

$$f_0 = \left(\frac{\gamma}{2\pi}\right)\sqrt{2H_E H_\parallel + H(H + H_D)}, \tag{1a}$$

$$f_1 = \left(\frac{\gamma}{2\pi}\right)\sqrt{2H_E H_\perp + H_D(H + H_D)}, \tag{1b}$$

where γ is the gyromagnetic ratio, $H$ is the applied dc magnetic field, $H_E$ is the exchange field, $H_\parallel$ ($H_\perp$) is the in-plane (out-of-plane) anisotropy field, and $H_D$ is the DMI field. The parameters agree well with previous reports (see SM Table1).

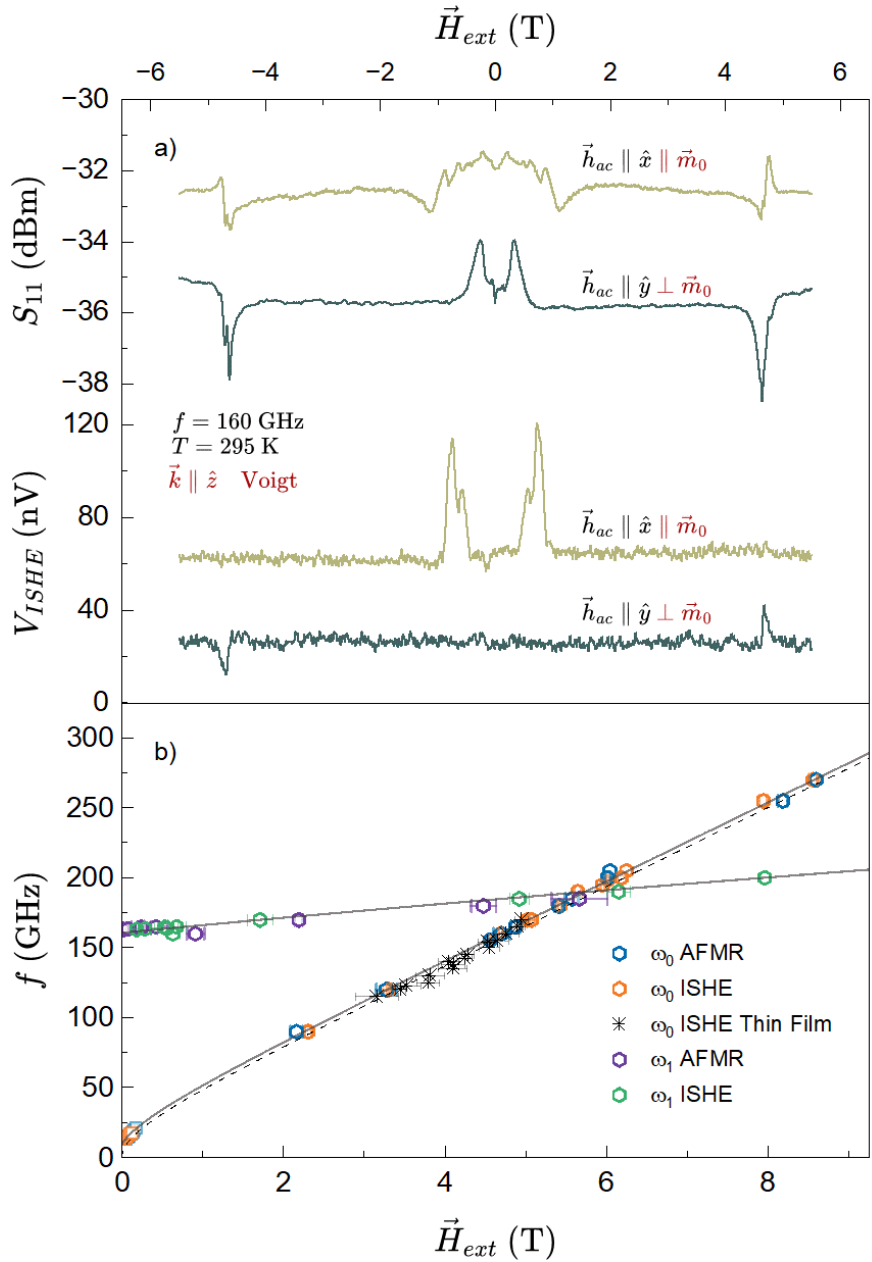

**Figure 2**: (a) AFMR and ISHE signals for 160 GHz at room temperature for linearly polarized microwaves propagating in Voigt geometry. Yellow curves select spin pumping from the optical mode, while blue curves select the acoustic mode. (b) Resonance dispersion of hematite for frequencies up to 255 GHz for absorptions in positive DC magnetic field, Solid lines indicate fittings by equations 1 and 2 to the bulk data sets, while the dashed line corresponds to the thin film data. Error bars are extracted from linewidth fittings.

In AFM systems, microwave absorption is governed by driving the magnetic moment $\vec{m}$ through the oscillating magnetic field $\vec{h}_{ac}$. In the acoustic mode, $\vec{h}_{ac}$ would be strictly collinear with $\vec{m}$ should the DMI be absent, which in our geometry means that they would both undergo linear oscillations along z. However, the DMI gives rise to a net equilibrium magnetization $\vec{m}_0$ along x, which renders the dynamics of $\vec{m}$ perpendicular to $\vec{m}_0$, forming an elliptical precession on the y-z plane[17,26]. Correspondingly, the selectivity requires that $\vec{h}_{ac} \perp \vec{m}_0$, so $\vec{h}_{ac}$ can be polarized along either y or z. On the other hand, the optical mode is characterized by a linearly vibrating $\vec{m}$ along its equilibrium direction (*i.e.*, $\vec{m} = \vec{m}_0 + \hat{x}\,\delta m \cos \omega t$) even in the presence of DMI so that $\vec{h}_{ac}$ must be polarized along x in order to satisfy the resonance condition. It should be noted that while the microwave delivers energy by directly driving $\vec{m}$, the ensuing spin pumping does not necessarily originate from $\vec{m}$; it could stem from the Néel vector $\vec{n}$ whose dynamics is intrinsically connected to $\vec{m}$[32]. This is a universal property of spin pumping in AFMs, not just the easy-plane systems. Basing on the above considerations, we have conducted experiments with the microwaves arranged in both the Voigt geometry, as shown in Fig. 2, and the Faraday geometry, as shown in Fig. 3. Comparing these two sets of measurements, we find that only when $\vec{h}_{ac}||x$ (realized exclusively in the Voigt geometry) could the optical mode resonance generate spin pumping (see lower traces of Fig. 2a), whereas neither $\vec{h}_{ac}||y$ nor $\vec{h}_{ac}||z$ can produce a similar signal. In particular, the Faraday geometry can never satisfy the $\vec{h}_{ac} || x$ condition, because $\vec{k}$, $\vec{H}$, and $\vec{m}_0$ are all aligned along $\hat{x}$. This is explicitly reflected in Fig. 3, where the spectroscopic signals (upper spectra) are clearly observed for the acoustic resonance (at higher fields), while the optical resonance peaks are much weaker than those in Fig. 2a. In sharp contrast to the Voigt geometry, the ISHE signals (lower traces) are only observed for the acoustic mode when $\vec{h}_{ac} \perp \vec{m}_0$, while absorptions near the optical resonance positions (at lower fields) are totally absent. At this point, it is instrumental to ponder on Ref. [21], where only the Faraday geometry is adopted, so the absence of spin pumping associated with the optical mode is just consistent with ours. In other words, a visible ISHE signal owing to the optical mode resonance is unique to the Voigt geometry, which, as shown in Fig. 2a, turns out to be substantially larger than the acoustic one. These results imply that spin-to-charge conversion highly depends on the microwave polarization and that a 90-degree rotation of the microwave polarization can act as a switch for the conversion process.

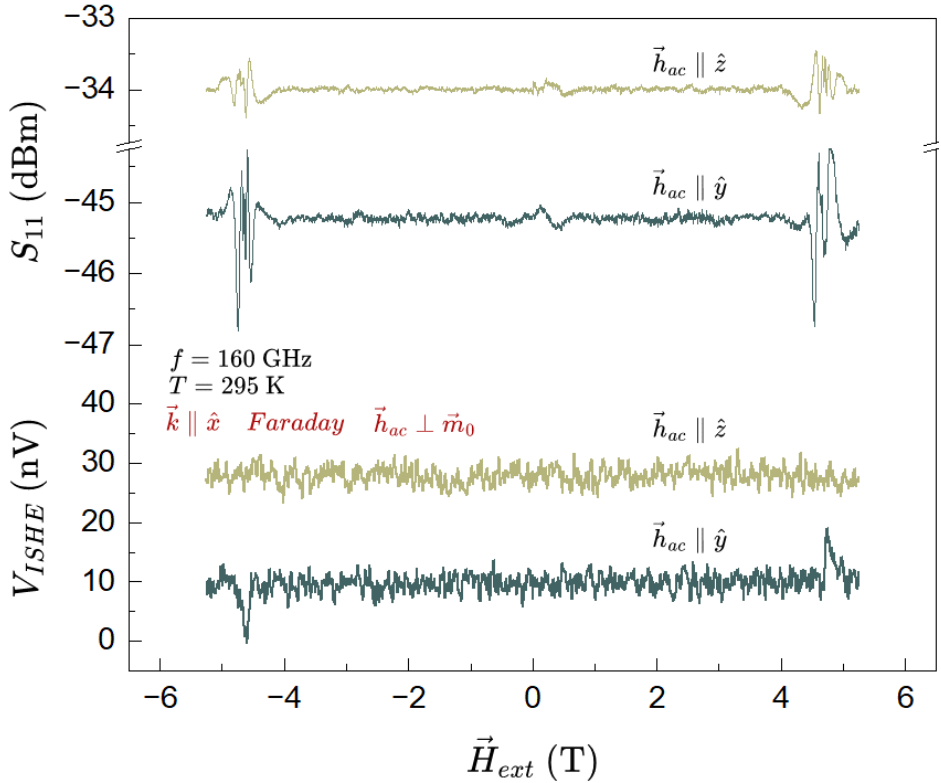

**Figure 3:** AFMR and ISHE signals for 160 GHz in ambient conditions for linearly polarized microwaves propagating in Faraday geometry. Yellow curves select spin pumping from the optical mode, while blue curves select the acoustic mode.

We note that the ISHE signals generated by the acoustic mode follow the expected sign reversal upon switching the direction of the applied magnetic field, which is consistent with previous studies[17,26]. On the other hand, the ISHE signals originating from the optical mode (unique to the Voigt geometry) appears to be invariant under field inversion, contradicting with the ordinary polarity-dependence of pumped spin-currents. This intriguing behavior calls for not only a detailed theoretical investigation but also further exploration allowing selectively tuning the microwave polarization between linear and circular modes, which will be published elsewhere [Fritjofson, G. et al., "Probing Polarization-Tunable Sub-Terahertz Spin Pumping in Bimodal α-$Fe_2O_3$/Pt", to be submitted to NPJ].

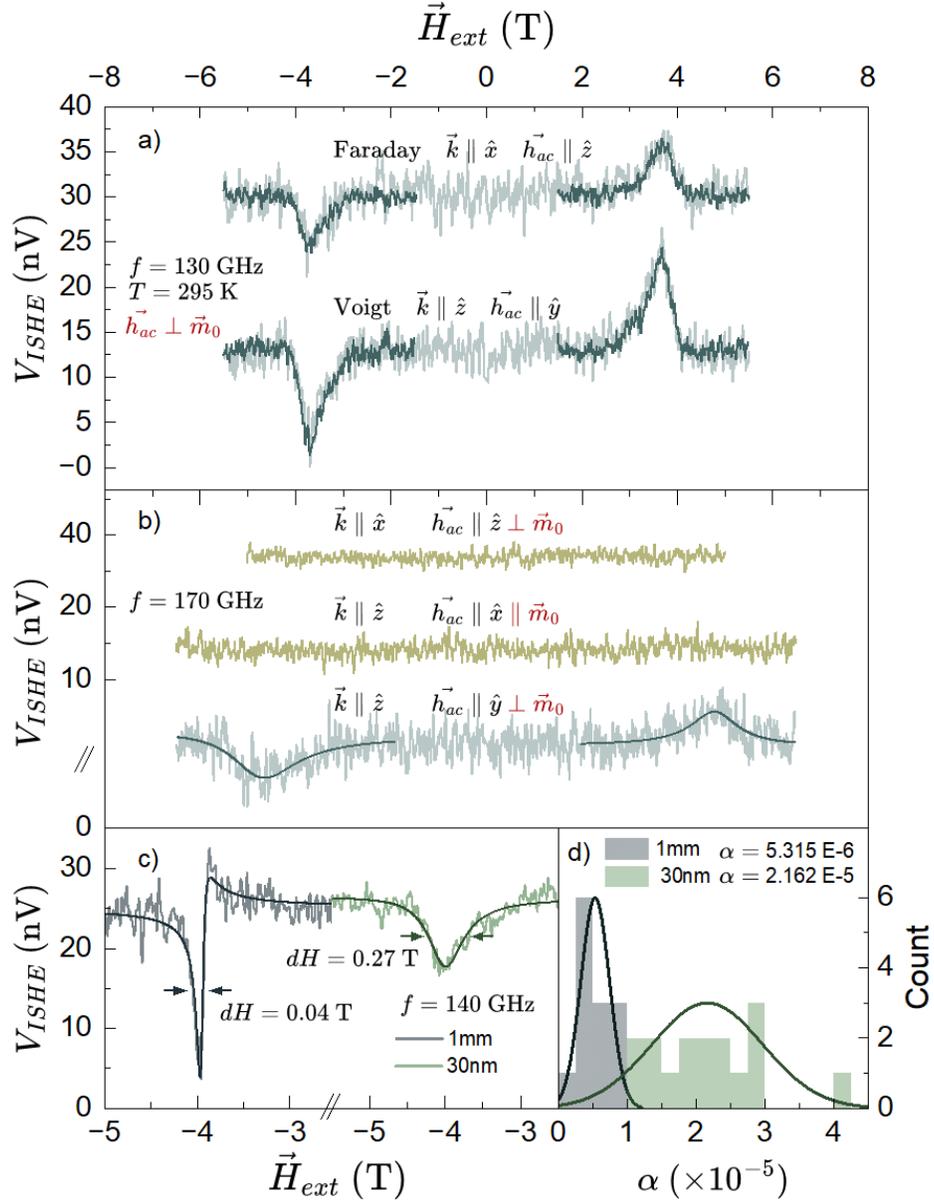

**Figure 4**: (a) ISHE signals for 130 GHz at ambient conditions for α-Fe2O3(30nm)/Pt(5nm) excited by $\vec{h}_{ac} \perp \vec{m}_0$ in Faraday and Voigt geometries. Dark curves are the result of averaging 4-5 field sweeps (b) ISHE signals for 130 GHz in Faraday and Voigt geometries. Yellow curves are chosen to select spin pumping from the optical mode and blue curves to select the acoustic mode. (c) ISHE signals for 140 GHz for 1mm and 30nm samples and corresponding linewidths. (d) Gilbert damping, $\alpha$, distributions for 1mm and 30nm samples extracted from acoustic ISHE signals.

As mentioned at the beginning, we have also performed equivalent measurements with a thin-film heterostructure composed of α-Fe$_2$O$_3$(30 nm)/Pt(5 nm). For the film, we again choose the (0001) orientation grown on Al$_2$O$_3$ substrates, matching the device configuration in Fig. 1a and guaranteeing the equilibrium magnetization and Néel vectors lie primarily within the sample plane. Due to the tiny amount of magnetic moments in thin films, our experimental setup is not sensitive enough to observe the AFMR spectroscopically. We therefore must rely on the ISHE voltages to guide our understanding. Resonant ISHE signals are observed for frequencies from 115-175 GHz (See spectra in Figure S3) and the corresponding peak positions are plotted in Figure 1c (stars), in close agreement with the observations in bulk samples. Acoustic ISHE signals obtained at 130 GHz for $\vec{h}_{ac} \perp \vec{m}_0$ reveal the same polarization dependence of the pumped spin current as in the bulk sample in both the Faraday and Voigt geometries (Figure 4a). In the Faraday geometry, the microwaves are incident from the side of the sample and a smaller percentage of the beam profile drives the resonance (less power), thus the Voigt signals are larger than the Faraday signals. This is indeed observed in both bulk and thin-film samples but clearly observed here by a close comparison of the traces in the same figure panel. The heights of the ISHE voltage peaks are noticeably lower for the thin film, while the linewidth ($dH$) is dramatically larger. Figure 4c compares this linewidth change between the 1mm and 30nm samples for 140 GHz, showing a roughly six times increase in $dH$ from the bulk to the film. For each frequency $dH$ is extracted from Lorentzian fittings (Figure S1), and the resultant Gilbert damping parameters create the distribution in Figure 4d. The film possesses the ultra-low damping expected in hematite; however, it exceeds the bulk by an order of magnitude, which is expected from the increased level of inhomogeneity in thin-film samples. Still the area under the peak is comparable between the two samples, illustrating that the spin pumping originates within the vicinity of the interface, meaning no significant difference between a thin-film and a bulk heterostructure.

Figure 4b shows how the film responds to microwaves at 170 GHz, a frequency chosen to sample the optical branch. While spin pumping from acoustic-mode resonances is generated as before for $\vec{h}_{ac} \perp \vec{m}_0$ (blue curve), the optical-mode resonance is not observed for either geometry (yellow curves). This behavior, which persists for all frequencies, is opposed to the observations in the bulk sample for $\vec{h}_{ac} \parallel \vec{x} \parallel \vec{m}_0$ and compels a discussion. We propose two possible explanations for this puzzling result. One possibility is that although a 30nm-thick Hematite film can efficiently develop coherent quasi-ferromagnetic resonance modes, as is also the case in thin ferro- and ferri-magnetic films, it may not be thick enough to support the coherent excitation of an optical magnetic resonance that requires a much larger out-of-plane oscillation of $\vec{n}$. As in the case of ferromagnets, there may exist a critical thickness for higher order modes to be efficiently excited. The amplitude and complexity of the spatial distribution of higher order magnon modes has been shown to diminish below a critical thickness in Py films[27]. Furthermore, in YIG it has been shown that the three-magnon splitting process which gives way to higher order standing spin waves is thickness dependent and prohibited for thin enough films[28]. This option may explain why pure AFM spin pumping has not been reported yet from a thin-film sample and would have important consequences that may prevent the use of pure AFM dynamical modes in nanoscale devices where thin films will be required. Another possibility is that the optical mode is somehow more sensitive to disorder than the acoustic mode and may be extinguished by extrinsic damping processes, either by the inhomogeneous broadening brought about by spatial variations in magnetic anisotropy across the film, or by a two-magnon scattering annihilation process for the higher order modes. The observed large resonance linewidths in films could be attributed to one or both of these processes[29–31]. By fitting the dispersion in the film, we obtain a very low value for the in-plane anisotropy $H_\parallel$, a three orders of magnitude reduction from the bulk. This heavily supports the case for diffused magnetic anisotropy as the origin for inhomogeneous broadening. The dramatic relaxation of the film lattice as compared to the bulk can be observed in the disappearance of the Morin transition ($T_m$), a strain-induced magnetic state change which appears in the bulk at 261K for our sample (Supplementary Figure S4a). Finally, an alternative explanation would rely on a substantial enhancement of the transverse anisotropy in thin-films hematite, moving the optical mode to higher frequencies than those sampled in this work (175GHz). Indeed, we observe a change in the in-plane anisotropy from our dispersion fittings. We believe that the elucidation of this open question calls for an active experimental and theoretical investigation of this effect for different film thicknesses.

## Data Availability

The datasets generated during and/or analyzed during the current study are available from the corresponding author upon reasonable request.

## Acknowledgements

This work was funded under AFOSR Grants FA9550-19-1-0307 and FA9559-24-1-0290. The work at Ohio State University (growth and characterization of α-Fe$_2$O$_3$ epitaxial films as well as purchase and characterization of α-Fe$_2$O$_3$ bulk crystals) was supported by the Department of Energy (DOE), Office of Science, Basic Energy Sciences, under Grant No. DE-SC0001304.